\documentclass{arxiv}
\usepackage{amssymb}
\usepackage{epsfig}
\usepackage{subfigure}
\usepackage{cite}
\usepackage{rotating}
\usepackage{caption}

\title{Citations to articles citing Benford's law: a Benford analysis}
\author{Tariq Ahmad Mir}
\address{
Nuclear Research Laboratory, Astrophysical Sciences Division, Bhabha Atomic Research Centre, Srinagar-190 006, Jammu and Kashmir, India. \\email:  taarik.mir@gmail.com\\}

\begin{document}

\catchline{}{}{}{}{}

\maketitle


\section{Abstract}
The occurrence of first significant digits of numbers in large data is often governed by a logarithmically decreasing distribution called Benford's law (BL), reported first by S. Newcomb (SN) and many decades later independently by F. Benford (FB). Due to its counter-intuitiveness the law was ignored for decades as a mere curious observation. However, an indication of its remarkable resurgence is the huge swell in the number of citations received by the papers of SN/FB. The law has come a long way, from obscurity to now being a regular subject of books, peer reviewed papers, patents, blogs and news. Here, we use Google Scholar (GS) to collect the data on the number of citations received by the articles citing the original paper of SN/FB and then investigate whether the leading digits of this citations data are distributed according to the law they discovered. We find that the citations data of literature on BL is in remarkable agreement with the predictions of the law. 
\section{Keywords}

Benford's law; citations; Google Scholar


\section{Introduction}	
The common perception that first digits of decimal numbers in large data are distributed uniformly, irrespective of their magnitude, is only superficial. In reality the distribution of digits in large tabulated data is contrary to intuition in that the digits 1 to 9  appear as first digits with varying proportions with digits of smaller magnitude appearing as the first digits of numbers more frequently than do the digits of larger magnitude. The phenomenon was first reported by SN in 1881 after he noticed that the initial pages of the logarithmic tables were dirtier than the last ones\cite{Newcomb}. It attained much popularity only after FB in 1938 independently rediscovered it through a similar observation. FB tested the accuracy of his observation by analyzing large data sets collected from diverse fields and established the law in the form of following empirical mathematical equation\cite{Benford}
\begin{equation}
P(d)= log_{10}(1+\frac{1}{d}), d= 1, 2, 3...,9
\end{equation}
where $P(d)$ is the probability of a number having the first non-zero digit d and $log_{10}$ is logarithmic to base 10. Thus in a given data set the theoretical proportions for each of the digits from 1 to 9 to be first significant digit are as shown in Table 1.
\begin{table}[h]
\tbl{The distribution of first significant digits as predicted by BL}
{\begin{tabular}{@{}ccccccccccc@{}} \toprule
Digit \hphantom{00} & 1 & 2 & 3 & 4 & 5 & 6 & 7 & 8 & 9 \\
Proportion \hphantom{00} & 0.301 & 0.176 & 0.125 & 0.097 & 0.079 & 0.067 & 0.058 & 0.051 & 0.046 \\ \botrule
\end{tabular} \label{ta1}}
\end{table} 

The mathematical form  of BL, equation (1), is quite simple. However, a comprehensive mathematical explanation of the law has proven to be equally difficult\cite{Berger}, though some of its distinguishing properties like scale invariance\cite{Pinkham} and base invariance\cite{Hill, Hill1} have been explained. Nevertheless, its presence has been confirmed for data from numerous phenomena\cite{Sandron, Mir, Mir1, Shao, Sambridge, Nigrini, Giles, Mir2, Mir3}. On the other hand, the universal nature of the the law is clearly ruled out by the presence of many data sets whose digit distributions deviate from the predictions of the law. There is no absolute criteria to explain why the law holds for data from such a variety of processes or breaks down for some others\cite{Durtschi}. Usually break down of BL is attributed to human thought process which influences or restricts the appearance of numbers in data\cite{Hill2}. However, recently through the analysis of large birth rate data, it has been pointed out that the break down of the law may also be due to natural reasons\cite{Ausloos}. Several variants of the law have been proposed to explain the distribution of digits deviant with respect to the law\cite{Clippe, Fu}. 

A recent addition to the ever expanding domain of BL research is the field of scientometrics. The frequencies of the first digits of the citation numbers of articles of individual authors have been found to be consistent with the expectations from BL\cite{Breuer}. Further, the validity of the law has been studied for the number of articles published, citations received and the impact factors of the scientific journals and only the data on citations received by the articles is found to follow BL\cite{Campanario}. For the journals indexed in the JCR Sciences and Social Sciences Editions from 2007 to 2011 the non-conformity of the data in case of number of published articles in journals, after categorizing them according to the country of their origin and the subjects they cover, was re-pointed out\cite{Alves}. The extended JCR data from 2007 to 2014 on bibliometric indicators like number of aricles, citations, impact factors, citation half-lives and immediacy index has been tested for compliance to BL\cite{Alves1}.

The informetric nature of BL has been explained after its derivation from the classical informetric law of Zipf with a tunable exponent $\beta$$>0$\cite{Pietronero, Egghe}. The use of this generalized form of BL with an optimized $\beta$, improves the compliance of the data of Ref.\cite{Campanario} significantly than that to the classical BL\cite{Egghe1}. 

The application of BL, quite straightforward owing to its simple mathematical form, to the plethora of high quality data readily available from internet, has over the years resulted in the  accumulation of large body of literature to the point that relevant research has grown into a full-fledged field. For the quick reference of the researchers attempts have been made to compile the list of publications on the different applications and mathematical aspects of the law. Hurliman\cite{Hurliman} gathered 349 papers on BL and another similar survey lists 613 publications\cite{Beebe}. Further, the largest single online resource updated continuously and exclusively dedicated to the publications on the different aspects of BL across multiple disciplines is ``Benford Online Bibliography''\cite{Online}. This data base as on September 21, 2015 had indexed 912 articles. Other than these compilations, the only scientometric/bibliometric analysis of Benford literature, so far, is Costa\cite{Costa} who, for the period of years, 1988 to 2011, collected 45 publications on the applications of the law in the specific field of accounting audit. The authors further tabulated the publications according to the country of their origin, number of publications per year, the journals in which the papers were published and the number of papers per author. 

In the backdrop of above mentioned scientometric studies of BL, one pertinent question coming to fore is that whether the pattern of digits predicted by the law is found within the citations data of its relevant literature. Inspired by the results of study of Refs.\cite{Campanario, Breuer} that citations data of the articles in general follow BL we here investigate whether same holds true for the data on the number of citations received by the articles citing the original papers of SN/FB. We use GS\cite{GS} to first collect the data on the number of citations received by all the articles that cite the two original papers where the discovery of BL was first reported. The occurrence distribution of the first significant digits in these citations data is then investigated.

\section{Data}
The two compilations\cite{Hurliman, Beebe} of Benford literature contain no information about the citations received in turn by the citing articles whereas \cite{Online} provides citations only for the items indexed within its own database. We use GS, the scholarly search tool of the world’s largest and most powerful search engine i.e. Google, to collect the citations data on BL.
GS searches for the articles related to a given article across the items indexed within a variety of data bases. It locates potentially relevant articles on a given subject by identifying subsequent articles that cite a previously published article. The  \textit{Cited by \#} feature of GS enables researchers to trace interconnections among authors citing articles on the same topic and to determine the frequency with which others cite a specific article\cite{Noruzi}. 

In order to look for the scholarly literature on BL we searched GS by entering directly the title of SN paper ``Note on the frequency of use of different digits in natural numbers'' and then separately of FB ``The law of anomalous numbers''. From the \textit{Cited by \#} function of the GS we first noted down the total number ``\#'' of citations received by each of the two papers. Further click on the \textit{Cited by \#} function leads to individual articles citing the original paper of SN/FB and we again noted down the number of citations in turn received by each of these citing articles. Within the search results of each paper we use continuous updation of GS and its \textit{custom range} function, to collect two types of citations data. 

The number citations to any given article on GS increases monotonically with time. This is due to the automatic updation of GS. The frequency of updation of GS is a business secret for Google\cite{Jasco} and to allow for an appreciable change in the citation numbers of SN/FB and of the articles linked to these two papers, by \textit{Cited by \#} function, we chose a considerable time lag of one and two years for our successive searches. The searches were carried out on September 30 of years 2012, 2013 and 2015 and corresponding citations to SN are 310, 410, 544 (Table 3) and respectivel to FB are 617, 748 and 1011 (Table 4).


The menu on the left side of the search results page of GS has various date filters for narrowing down the search results according to ones specific aims. These filters allow the selection of search results from any time, sorting the results by date and since the year of publication of articles in recent years such as “since 2012,” “since 2015,” etc. For our purpose we use the \textit{custom range} function to select a time range of our choice. A click on the “custom range” link creates two blank fields for entering the dates in years to bring up the articles published within those years only. 

For SN we entered 1881, year of its publication, in the first field of the custom range function and 2006 in the second field. The search for 1881-2006 period turned up 139 articles citing SN. We fixed the year in first field at 1881 and successively incremented the year in the second field to collect the number of records up to the year 2016. The search for years below 2006 resulted in number of records which is too small for any meaningful Benford analysis and hence are not considered for the said purpose. The citations data corresponding to SN was collected by accessing the GS on February 25, 2016. 

We start our search for FB with the year 1938 in the first field and 1999 in the second field of the custom range function. The search for period of 1938-1999 showed a total of 103 citations. In this case the search for years before 1999 yields samples of smaller size and hence are not shown. The analysis of search results from 1938 upto year 2016 is shown in Table 6. This data was collected by accessing the GS on February 26, 2016. 

\subsection{Analysis and Results}
A GS search carried out on September 30, 2013 showed a total of 748 citations to FB. After inspecting the citation numbers of each article it was found that 299 articles in turn were uncited i.e. have zero citations. The remaining 449 articles have non-zero citations and we noted down the number of citations in turn received by each of these articles. Thus number of records in the data set corresponding to FB for September 30, 2013 is 449. On the same date SN had 410 citations out of which 179 in turn were uncited and 231 had non-zero citations. We again noted down the citation numbers of each of these 231 articles and hence the sample size corresponding to SN is 231. For these monthly data sets we identified the first digits of all the records. For example, on September 30, 2013, the number of citations to Pietronero (2001)\cite{Pietronero} is 118 and this has a first digit 1 and Raimi (1976)\cite{Raimi} is cited 234 times and this has a first digit 2. Similarly, Hill\cite{Hill} is cited 342 times and this has a first digit 3.

The distribution of first digits of the September 30, 2013 citations data for both FB and SN are shown in Table 2. For FB, $N_{Obs}$, the number of times each of the digits from 1 to 9 (column 1) appears in the citations data as first significant digit is shown in columns 4 of Table 2. We also show (column 5) $N_{Ben}$, the corresponding counts for each digit as predicted by BL: 
\begin{equation}
N_{Ben}= N log_{10}(1+\frac{1}{d})
\end{equation} 
along with the root mean square error ($\Delta{N}$) calculated from the binomial distribution 
\begin{equation}
\Delta{N}= \sqrt{NP(d)(1-P(d))}
\end{equation} 
where $N$ is the total number of records in the data.

From a visual inspection of the Table 2 it is found that the observed and expected digit distributions are in reasonable agreement. This is further illustrated in sub figures (a) and (b) of Fig. 1 where the observed proportion of the first digits with those expected from BL are compared. We also show the equal proportion of 11\% for all the digits as would usually be expected from the uniform distribution. In both cases a clear deviation from the uniform distribution is seen and pattern of the decrease in proportion of the digits with increase in magnitude of the respective digits as predicted by BL can be easily seen. However, due to the finite size of the samples under investigation it is impossible to have a distribution of digits exactly same as a Benford one\cite{Nigrini1}. This necessitates the quantification of the level of agreement of the observed and the theoretical digit distributions. We use Pearson's $\chi^{2}$ test to quantify the goodness-of-fit between two distributions
\begin{equation}
 \chi^{2}(n-1) =\sum_{i=1}^n\dfrac{(N_{Obs}-N_{Ben})^{2}}{N_{Ben}}
\end{equation}

\begin{table}[h]

\tbl{The significant digit distribution of citations received by the articles citing SN and FB on September 30, 2013}
{\begin{tabular}{@{}llllll@{}} \toprule
First Digit & SN & $N_{Ben}\pm\Delta{N}$ & FB & $N_{Ben}\pm\Delta{N}$ \\
& N=231 &  &N=449  &  &  \\ \colrule
$1$ \hphantom{00} & 78 & 69.5$\pm$7.0 & 140  & 135.2$\pm$9.7  \\
$2$ \hphantom{00} & 53 & 40.7$\pm$5.8  & 90   &  79.1$\pm$8.1 \\
$3$ \hphantom{00} & 29 & 28.9$\pm$5.0  & 54   & 56.1$\pm$7.0 \\
$4$ \hphantom{00} & 22 & 22.4$\pm$4.5  & 47   & 43.5$\pm$6.3 \\ 
$5$ \hphantom{00} & 15 & 18.3$\pm$4.1  & 33   & 35.5$\pm$5.7 \\
$6$ \hphantom{00} & 11 & 15.5$\pm$3.8  & 27   & 30.1$\pm$5.3 \\
$7$ \hphantom{00} & 11 & 13.4$\pm$3.5  & 28   & 26$\pm$4.9 \\
$8$ \hphantom{00} & 6 & 11.8$\pm$3.3   & 13   & 23$\pm$4.7 \\
$9$ \hphantom{00} & 6 & 10.6$\pm$3.2   & 17   & 20.5$\pm$4.4 \\ 
\botrule
\textbf{Pearson's} $\chi^{2}$ \hphantom{00} & \bf11.919  &   &  \bf7.623 &   & \\ \botrule
\end{tabular} \label{ta1}}
\end{table} 

\begin{table}[]
\tbl{The GS citations data of the articles citing SN} 
{\begin{tabular}{@{}ccccccc@{}} \toprule
Date & Total \#   &  \# of  & (N), \# of  & $\chi^{2}$  &  Compliance to&\\ 
 & of  & uncited & cited &   & BL &\\
&  citations  & citations & citations &    &  &\\ \colrule
September 30, 2012 & 342 &  146\hphantom{00}  &  196 &  14.531 & Yes & \\
September 30, 2013 & 410 &  179\hphantom{00}  &  231 &  11.919 &  Yes &\\
September 30, 2015 & 544 &  226\hphantom{00}  &  318 &  7.968 &  Yes &\\
\botrule
\end{tabular} \label{ta1}}
\end{table} 

\begin{table}[]
\tbl{The GS citations data of the articles citing FB}
{\begin{tabular}{@{}ccccccc@{}} \toprule
Date & Total \#   & \# of   & (N), \# of  &  $\chi^{2}$  & Compliance to&\\ 
  &  of  & uncited  & cited  &  & BL &\\ 
&  citations  & citations & citations &   & &\\ \colrule
September 30, 2012 & 617 & 250 \hphantom{00} & 367 & 14.891 & Yes &\\
September 30, 2013 & 748 & 299 \hphantom{00}  & 449 & 7.623 & Yes &\\
September 30, 2015 & 1011 & 412\hphantom{00}  & 599 & 9.311 & Yes &\\
\botrule
\end{tabular} \label{ta1}}
\end{table}

\begin{table}[h]
\tbl{The significant digit distribution of citations received by the articles citing (i) SN from 1881-2015 and (ii) FB from 1938-2015}
{\begin{tabular}{@{}llllll@{}} \toprule
First Digit & SN & $N_{Ben}\pm\Delta{N}$ & FB & $N_{Ben}\pm\Delta{N}$ \\
& N=333 &  & N=635 &  &  \\ \colrule
$1$ \hphantom{00} & 112 & 100.2$\pm$8.4 & 206  & 191.2$\pm$11.6\\
$2$ \hphantom{00} & 59  & 58.6$\pm$6.9  & 104  & 111.8$\pm$9.6\\
$3$ \hphantom{00} & 36 & 41.6$\pm$6.0   & 76   & 79.3$\pm$8.3\\
$4$ \hphantom{00} & 39 & 32.3$\pm$5.4   & 69   & 61.5$\pm$7.4\\
$5$ \hphantom{00} & 24 & 26.4$\pm$4.9   & 48   & 50.3$\pm$6.8\\
$6$ \hphantom{00} & 18 & 22.3$\pm$4.6   & 39   & 42.5$\pm$6.3\\
$7$ \hphantom{00} & 23 & 19.3$\pm$4.3   & 44   & 36.8$\pm$5.9\\
$8$ \hphantom{00} & 13 & 17.0$\pm$4.0   & 27   & 32.5$\pm$5.5\\
$9$ \hphantom{00} & 9 & 15.2$\pm$3.8    & 22  & 29.1$\pm$5.3\\ 
\botrule
\textbf{Pearson's} $\chi^{2}$ \hphantom{00} & \bf8.792 &   &  \bf7.176 &   & \\ \botrule
\end{tabular} \label{ta1}}
\end{table} 
For a data set with $n-1=9-1=8$ degrees of freedom, the critical value of $\chi^{2}$ for the acceptance of null hypothesis at $5\%$ level of significance is $15.507$. From the last row of the Table 2 the Pearson's $\chi^{2}$ for both data sets turn out to be less than the critical value and therefore null hypothesis is accepted and we conclude that the citations data corresponding to both SN and FB collected on September 30, 2013 follow BL.

\begin{figure}
\begin{center}
\begin{minipage}[b]{.9\linewidth}
\vspace*{-5pt}
\hspace*{5pt}
\centering
\begin{tabular}{cc}
\hspace*{-55pt}
\vspace*{-70pt}
\subfigure[Newcomb (1881)]{\label{fig:edge-a}\includegraphics[width=0.45\linewidth, height=0.55\linewidth, angle=270]{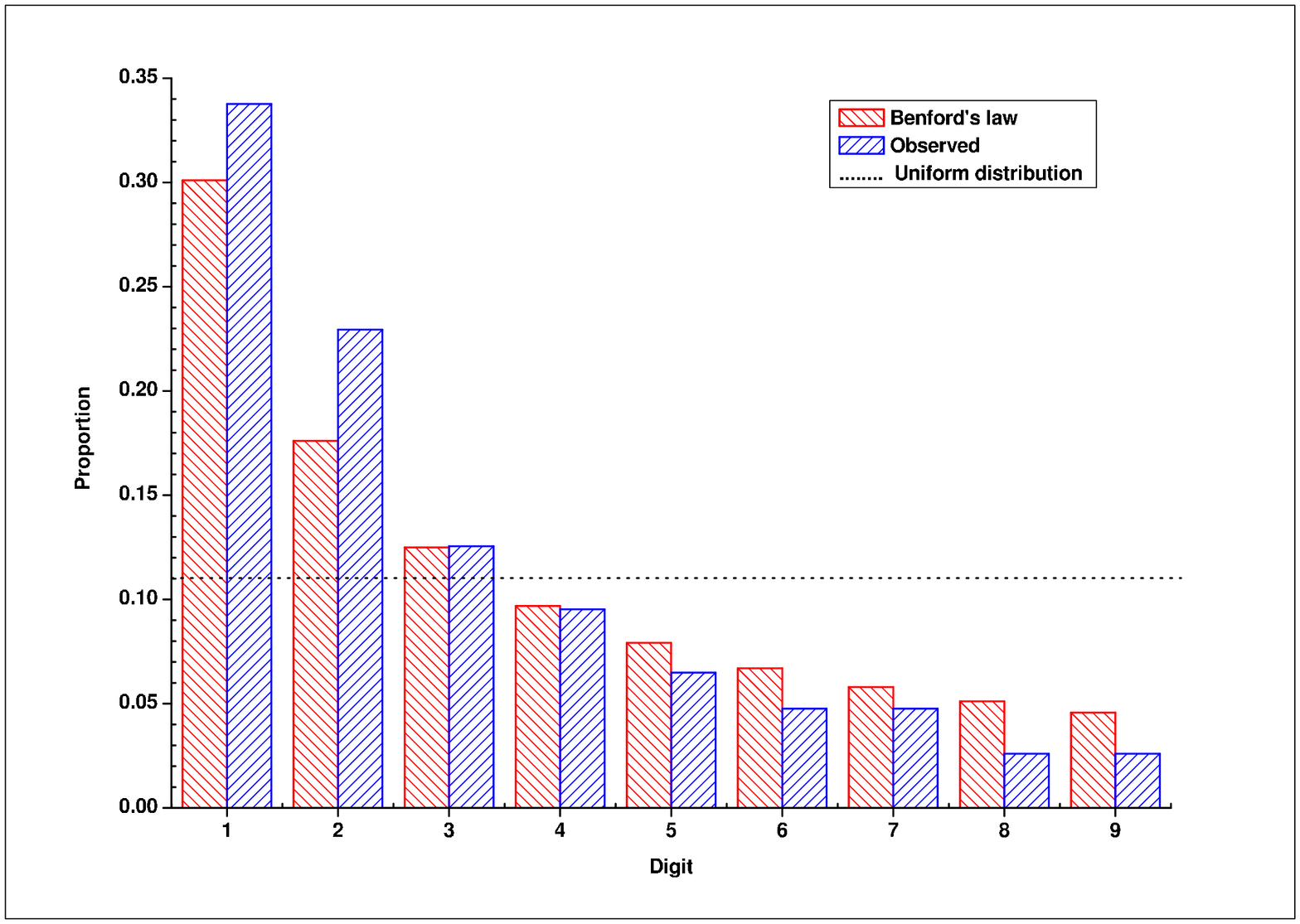}}
\hspace*{-5pt}
\subfigure[Benford (1938)]{\label{fig:edge-b}\includegraphics[width=0.45\linewidth, height=0.55\linewidth,  angle=270]{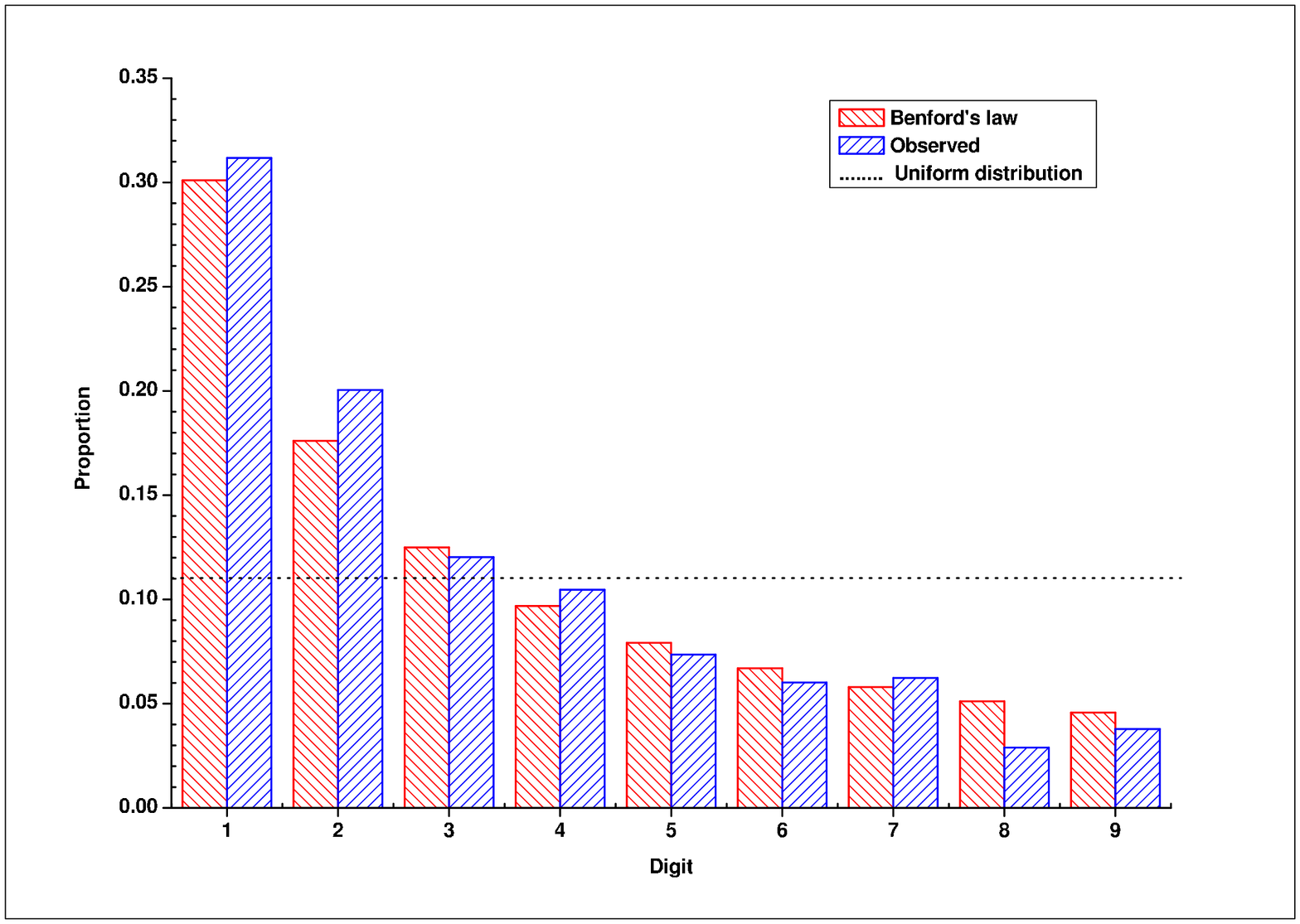}}
\hspace*{-74pt}
\vspace*{20pt}
\end{tabular}
\vspace*{50pt}
\hspace*{20pt}
\end{minipage}
\end{center}
\caption{The observed and theoretical proportions of first digits of citations received by the articles citing (a) Newcomb (1881) and (b) Benford (1938) on September 30, 2013. For comparison the proportions expected from uniform distributions are also shown.}
\label{fig:edge}
\end{figure}

\begin{figure}
\begin{center}
\begin{minipage}[b]{.9\linewidth}
\vspace*{-5pt}
\hspace*{5pt}
\centering
\begin{tabular}{cc}
\hspace*{-55pt}
\vspace*{-70pt}
\subfigure[Newcomb (1881)]{\label{fig:edge-a}\includegraphics[width=0.45\linewidth, height=0.55\linewidth, angle=270]{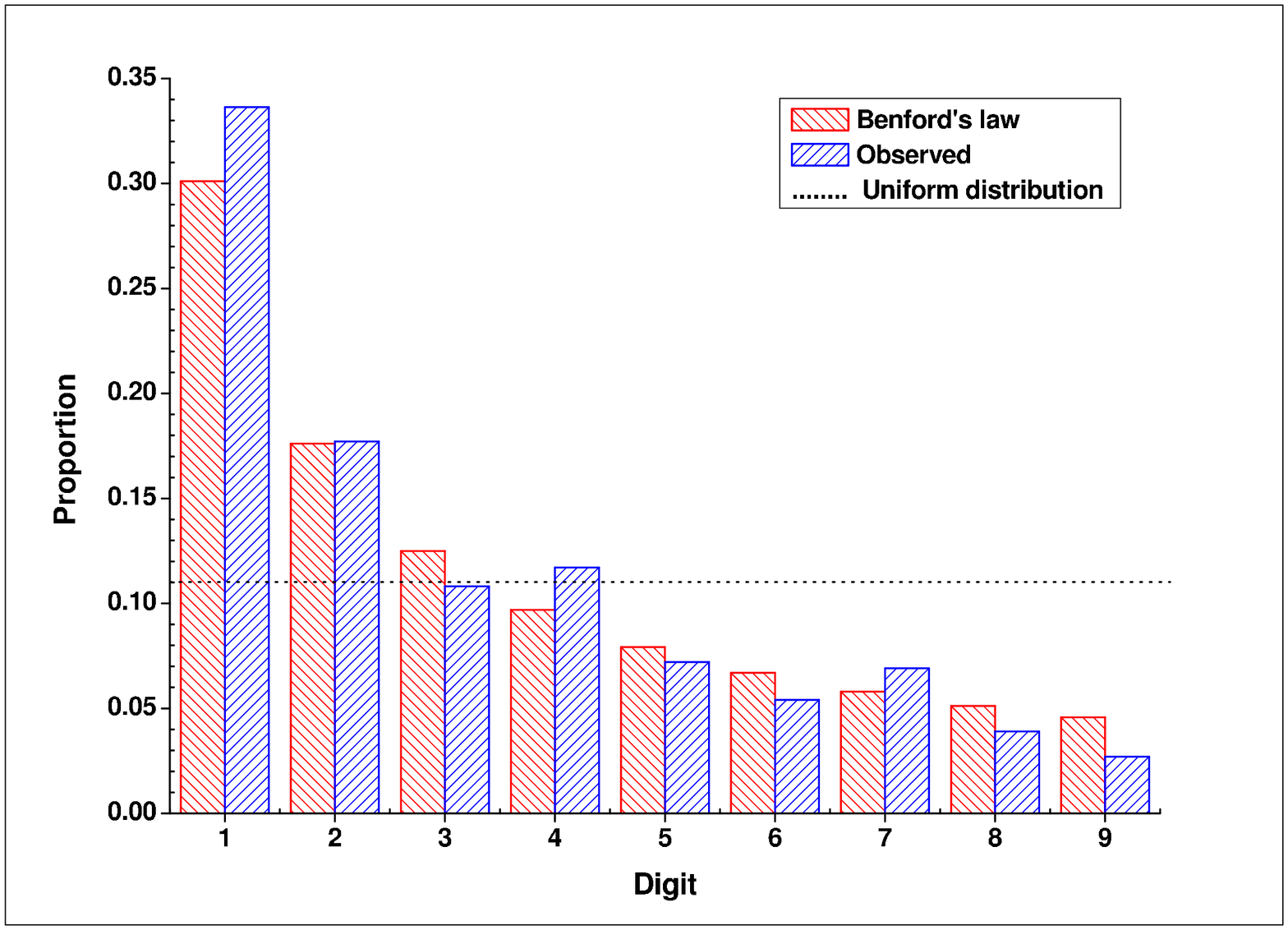}}
\hspace*{-5pt}
\subfigure[Benford (1938)]{\label{fig:edge-b}\includegraphics[width=0.45\linewidth, height=0.55\linewidth,  angle=270]{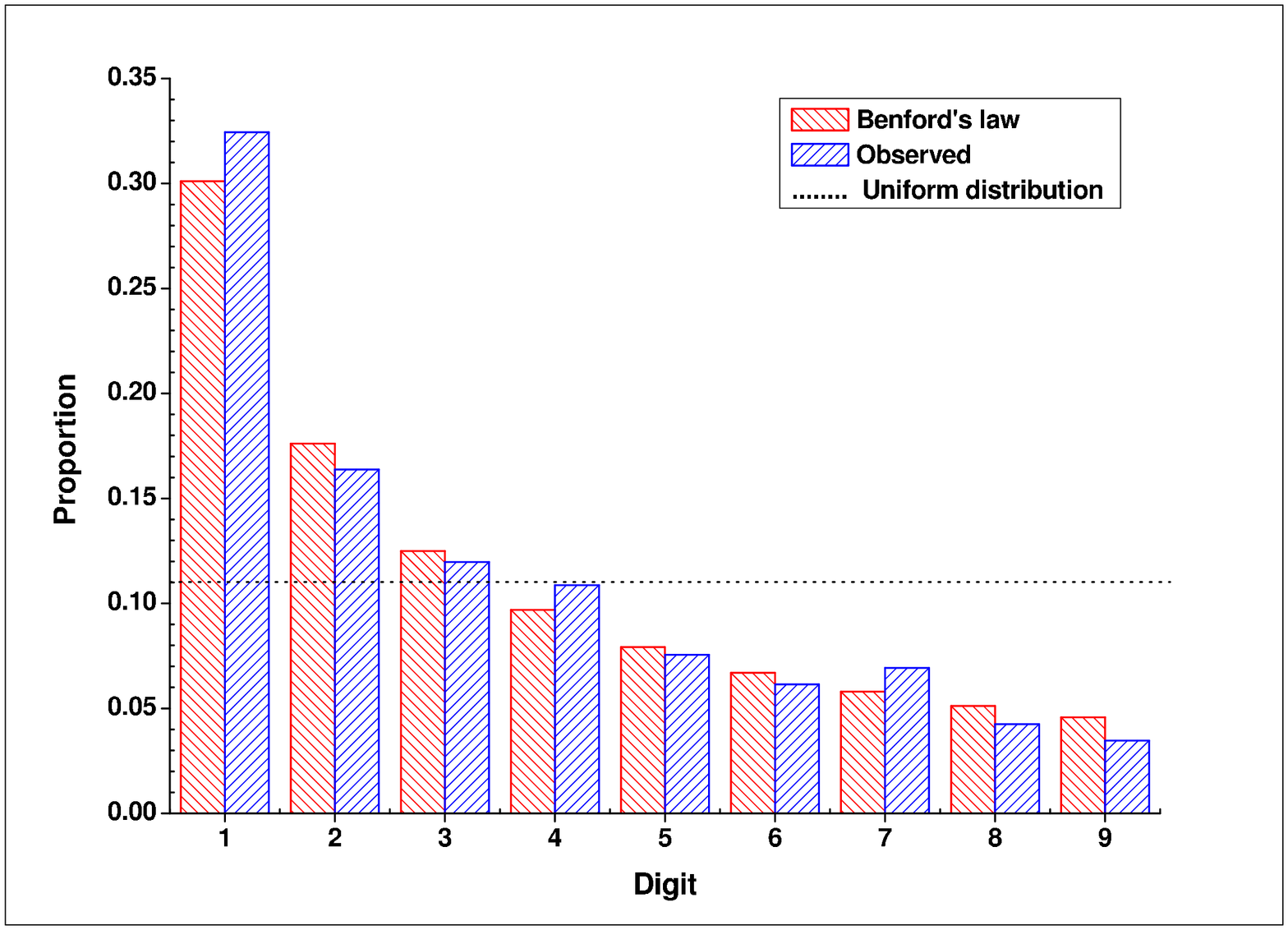}}
\hspace*{-74pt}
\vspace*{20pt}
\end{tabular}
\vspace*{50pt}
\hspace*{20pt}
\end{minipage}
\end{center}
\caption{The observed and theoretical proportions of first digits of citations received by the articles citing (a) Newcomb (1881) from 1881-2015 and (b) Benford (1938) from 1938-2015. For comparison the proportions expected from uniform distributions are also shown.}
\label{fig:edge}
\end{figure}
The analysis explained in Table 2 was performed for all the citations data sets collected by using two different features of GS as described above. We report the Pearson's $\chi^{2}$ for all the yearly citations data sets received by the articles citing SN/FB, collected using the timely updates of GS, in Tables 3 and 4 respectively. The total number of citations received by SN/FB on a given date of search (column 1) are shown in column 2. The number of citations which in turn are uncited are shown in column 3 whereas numbers of those which have received at least one or more citations, representing the size of sample for analysis, are indicated as N in column 4. The calculated $\chi^{2}$ (Column 5) for citations data sets are found to be less than the critical value of $15.507$. The results on compliance to BL are shown in column 6.

\begin{table}[]
\tbl{The GS yearly citations data of the articles citing SN}
{\begin{tabular}{@{}ccccccc@{}} \toprule
Year & Total \#   & \# of   & (N), \# of  &  $\chi^{2}$   & Compliance to&\\ 
  &  of  & uncited  & cited  & & BL  &\\ 
&  citations  & citations & citations &   &  &\\ \colrule
2006 & 129 & 39 \hphantom{00} &  90 & 8.563 & Yes &\\
2007 & 151 & 42 \hphantom{00} &  109 & 9.815 & Yes &\\
2008 & 183 & 47 \hphantom{00} &  136 & 5.812 & Yes &\\
2009 & 232 & 63 \hphantom{00} & 169 &  5.339 & Yes &\\
2010 & 278 & 81 \hphantom{00} & 197 &  5.058 & Yes & \\
2011 & 331 & 110 \hphantom{00}  & 229 & 5.798 & Yes &\\
2012 & 393 & 130 \hphantom{00}  & 263 & 4.068 & Yes &\\
2013 & 437 & 152 \hphantom{00}  & 285 & 5.572 & Yes &\\
2014 & 498 & 177 \hphantom{00}  & 321 & 7.417 & Yes &\\
2015 & 558 & 225 \hphantom{00}  & 333 & 8.792 & Yes &\\
2016 & 567 & 231 \hphantom{00}  & 336  & 9.636 & Yes &\\
\botrule
\end{tabular} \label{ta1}}
\end{table} 

\begin{table}[]
\tbl{The GS yearly citations data of the articles citing FB}
{\begin{tabular}{@{}ccccccc@{}} \toprule
Year & Total \#   & \# of   & (N), \# of  & $\chi^{2}$   & Compliance to&\\ 
  &  of  & uncited  & cited  &  &  BL &\\ 
&  citations  & citations & citations &   &  &\\ \colrule
1999 & 103 & 13 \hphantom{00}  & 90 & 9.107 & Yes &\\
2000 & 111 & 15 \hphantom{00}   & 96 & 10.295 & Yes &\\
2001 & 131 & 22 \hphantom{00}  & 109 & 9.235 &  Yes &\\
2002 & 150 & 30 \hphantom{00}  & 120 & 7.057 & Yes &\\
2003 & 172 & 36 \hphantom{00}  & 136 & 4.976  & Yes &\\
2004 & 206 & 48 \hphantom{00} &  158 & 5.136 & Yes &\\
2005 & 236 & 53 \hphantom{00} &  183 & 6.322 & Yes &\\
2006 & 274 & 66 \hphantom{00} &  208 & 7.196 & Yes &\\
2007 & 321 & 80 \hphantom{00} &  241 & 9.945 & Yes &\\
2008 & 381 & 94 \hphantom{00} &  287 & 4.685 & Yes &\\
2009 & 451 & 114 \hphantom{00} &  337 & 3.438 & Yes &\\
2010 & 520 & 139 \hphantom{00} &  381 & 3.194 & Yes & \\
2011 & 616 & 179 \hphantom{00}  & 437 & 4.37 & Yes &\\
2012 & 719 & 219 \hphantom{00}  & 500 & 2.163 & Yes &\\
2013 & 813 & 265 \hphantom{00}  & 548 & 5.549 & Yes &\\
2014 & 931 & 322 \hphantom{00}  & 609 & 6.5 & Yes &\\
2015 & 1040 & 405 \hphantom{00}  & 635 & 7.176 & Yes &\\
2016 & 1060 & 421 \hphantom{00}  & 639  & 7.904 & Yes &\\
\botrule
\end{tabular} \label{ta1}}
\end{table} 

To explain the results of our analysis on the type of citations data collected using the \textit{custom range} function of GS we show in Table 5 the observed and expected digit distributions of citation numbers of articles citing SN from 1881 to 2015 and of articles citing FB form 1938 to 2015. The corresponding graphical representation is shown in sub figures (a) and (b) of Fig. 2. The analysis for all the yearly periods for SN and FB are shown in Tables 6 and 7. 

It may be noted from the data tables that the number of records for analysis increases over time for both types of citation data. BL works better for samples of larger size is well known because an increase in the number of records also increases the chances for digits to manifest themselves. Further, another rule of thumb for BL to hold is that data must span over many orders of magnitude. An article must have at least one citation to be considered for Benford analysis. This sets the lower limit for our data at digit 1. Amongst the results of the search done on September 30, 2015 there are six citations to SN/FB which in turn have 4-digit citation numbers. However, the article with 4-digit citations through out the period of search is Rothaus (1976)\cite{Rothaus}. Hence the data in the present analysis spans over three orders of magnitude. 

\section{Discussion}
The literature on BL is swelling as its importance and applications are being realized. Publications on BL have been enumerated\cite{Hurliman, Beebe, Online} and some studies have tested the accuracy of the law on scientometric data\cite{Breuer, Campanario, Alves, Alves1, Egghe, Egghe1} but there is none concerning its validity for the citations data of its own literature. To fill the gap we investigated the digit distributions of the citation numbers received by the articles citing the original papers of SN/FB with respect to the law they discovered. We found the compliance of citations data of the articles citing FB and SN to the law to be excellent. However, the conformity of the citation data for a given year may only be an isolated coincidence. To ascertain that our results are consistent and to check that they are unaffected by the variation (i) in the number of records and (ii) in the pattern of occurrence of first digits we collected citation numbers using two different features of the GS. 

The citations to SN/FB increases monotonically over time whenever GS is updated. Further, the number of citations to many individual articles linked to these two papers by \textit{Cited by \#} function also varies which in turn may change the first digit of the citation numbers of the articles. If an article with a certain number of citations appears in the search results on September 30 of 2012 it will reappear in the results of search done on September 30 of years 2013 and 2015 but its citation count may have changed during this period. For example, a search on September 30, 2012 showed 97 citations to Pietronero (2001)\cite{Pietronero} which increased to 144 when the search was repeated on September 30, 2015. For the same period the number of citations to Hill (1995)\cite{Hill} increased from 302 to 467. The first digit changed from 9 to 1 in the former case whereas for the latter it changed from 3 to 4. Much more prone to such digit variations are 1-digit and 2-digit citations. Thus variability in the digit distribution of this data set is due to (i) yearly increase in the number of records and (ii) variation in the citation numbers of individual articles over time.

In the citations data collected using the \textit{custom range} function the number of articles citing SN increases from 129 for the period of 1881-1999 to 567 for 1881-2016 in turn giving an increase in the number of records i.e. articles with non-zero citations, from 90 to 336 for respective periods. Similarly for FB the number of citations varies from 103 for 1938-1999 to 1060 for 1938-2016 thereby changing the number of records from 90 to 639. However, there is no change in the citation numbers of individual articles appearing in the search results of a given year as the search is expanded in steps of one year from the year of publication. This is due the fact that the data upto all the subsequent years from year of publications is accessed on the same date. Thus an article published upto a certain year, say 1999, will be included in the search results of all the succeeding years but its citation count will not change. For example, the number of citations to Hill (1995)\cite{Hill} stands at 501 for the entire custom range of years 1938-1999 to 1938-2016 and similarly for Pinkham (1961)\cite{Pinkham} citations remain at constant value of 241. Thus in this case the variability in the data is only due to increase in the number of records i.e. citation numbers of new articles. The compliance to BL is unaffected by such digit variations of the two types of citations data described above. 

The results of the present study are susceptible to the known inadequacies of GS like i) its failure to distinguish between the different origins of citations as it indexes documents, with questionable scholarly quality, other than peer-reviewed articles such as conference papers, books, book chapters, thesis and dissertations, ii) indexing of all kind of ghost citations and iii)  of the duplicate content\cite{Jasco, Jasco1}. Filtering out each of these factors, a sort of noise, for all the articles citing the FB/SN is too difficult a task to be attempted here. Thus a similar study with more precise data from Web of Science (WoS), considered to be highly authentic citation indexing service, is desirable. Currently the author does not have access to the WoS.
\section{Conclusion}
In the absence of any comprehensive mathematical explanation the recourse for resolving the mysteries of BL is in testing data from new phenomena since every process has its unique mechanism of generating data thereby giving it (data) specific characteristics. In the same line of thought to test BL on a new data set, we used GS to first collect the data on the number of citations received by the articles citing their original papers, reporting the law, and then studied the distribution of first digits within this data. The two types of yearly citations data corresponding to SN/FB are consistently in excellent agreement with the predictions of the law. 
 


\end{document}